# Adaptive Lookup for Unstructured Peer-to-Peer Overlays

K Haribabu, Dayakar Reddy, Chittaranjan Hota
Computer Science & Information Systems
Birla Institute of Technology & Science
Pilani, Rajasthan, 333031, INDIA
{khari, f2005462, c_hota}@bits-pilani.ac.in

Antii Ylä-Jääski, Sasu Tarkoma
Telecommunication Software and Multimedia Laboratory
Helsinki University of Technology
TKK, P.O. Box 5400, Helsinki, FINLAND
{antii.yla-jaaski@tml.hut, Sasu.Tarkoma@cs.helsinki}.fi

*Abstract*— **Scalability and efficient global search in unstructured peer-to-peer overlays have been extensively studied in the literature. The global search comes at the expense of local interactions between peers. Most of the unstructured peer-to-peer overlays do not provide any performance guarantee. In this work we propose a novel Quality of Service enabled lookup for unstructured peer-to-peer overlays that will allow the user's query to traverse only those overlay links which satisfy the given constraints. Additionally, it also improves the scalability by judiciously using the overlay resources. Our approach selectively forwards the queries using QoS metrics like latency, bandwidth, and overlay link status so as to ensure improved performance in a scenario where the degree of peer joins and leaves are high. User is given only those results which can be downloaded with the given constraints. Also, the protocol aims at minimizing the message overhead over the overlay network.**

## I. Introduction

Peer-to-peer (P2P) overlay networks are widely used as public file sharing networks. Data sharing P2P systems are capable of sharing huge amounts of data. For example, in April 2003 the KaZaA [5] P2P data sharing system reported over 4.5 million users sharing a total of 7 petabytes of data. Such a huge collection of data will be unusable without efficient lookup of the object being looked for.

P2P overlay networks are application-level logical networks built on top of the Internet. These networks maintain routing tables to enable efficient search and data exchange between peers. They don't require any special administrative or financial arrangement. They are self-organizing and adaptive, distributed and decentralized. They can support the distribution of storage and computational problems. P2P overlay networks are categorized as unstructured and structured [1]. An unstructured P2P system is composed of peers joining the network with some loose rules, without any prior knowledge of the topology. Freenet [2], Gnutella [3], FastTrack [4], and KaZaA [5] are examples of unstructured P2P overlay networks. These networks are typically power law networks (or scale free networks). Gnutella [3] is a traditional example of a power law network, where search has a high cost due to many connections between peers. In structured P2P overlay networks, network topology is tightly controlled and content is placed not at random peers but at specified locations that will make subsequent queries more efficient. Most of the structured P2P overlays are Distributed Hash Table (DHT) based. Content Addressable Network (CAN) [6], Tapestry [7], Chord [8], Pastry [9], Kademlia [10] and Viceroy [11] are some examples of structured P2P overlay networks.

In unstructured P2P network, lookup is based on forwarding the queries [12]. At each node the query is forwarded to neighbors. Unless the peer finds the item or the hop count of the query reaches zero, query is forwarded to neighbors. In this lookup approach, there are variations on how the query forwarding can be controlled without decreasing the chance of finding an item. The query forwarding is controlled by selectively choosing the neighbors. The selection is based on the information stored at the peer about its neighbors. The information is either past history or the indexes of the content available of neighbors. The controlled forwarding also happens by randomly selecting the neighbors which reduces the chance of finding an item.

Unstructured peer-to-peer overlay networks mostly consist of nodes which are home PCs. They are connected to network by a weak bandwidth connection. In this paper we present an approach to give freedom to the user to specify the constraints that should be satisfied for the results obtained. The results are ranked using a composite function that is expressed as a function of QoS metrics defined in the later sections. The result with the highest rank will be from the node that can satisfy the users constraints to the maximum.

The type of Quality of Service (QoS) introduces several factors that need to be taken into account. In this paper, we consider two parameters, bandwidth and link latency, at the link level. We consider one parameter past response or past interactions with the peer as the node level constraint. This paper presents a scalable and adaptive lookup approach that takes user preferences into account in choosing the best overlay route to fetch the object among the multiple locations.

## II. Related Work

The lookup problem in a P2P overlay refers to finding any given data item in a scalable manner. More specifically, given a data item stored at some dynamic set of nodes in the overlay, we need to locate it [13]. The unstructured overlays commonly use flooding [3], random walks [14], iterative deepening search [15], directed breadth first search (BFS) [15] to lookup content stored on other overlay peers.

Freenet [2] uses a symmetric lookup search where queries are forwarded from node to node based on the routing table entries that are built-up dynamically. It ensures anonymity by not forming any predictable topology and also by not associating an object with any server. Because of anonymity, search for an object needs to visit large fraction of nodes that is time consuming. In flooding technique [3], the query is forwarded to all the neighbors. To improve the scalability, it uses small time-to-live (TTL) counters. Though it reduces network traffic and load on peers, it also reduces the chances of finding a match. In k-walker random walks [14] the query is forwarded to k randomly selected neighbors. Those neighbors in turn forward to k randomly selected neighbors. Although this search method reduces the network load but massively increases the search latency. In iterative deepening search [15], consecutive BFS at increasing depths is performed to locate an object in the P2P overlay. This search method also increases network load and duplicate query messages. In this technique, at every node the query is forwarded to all neighbors except the one who sent the query. In GUESS (Gnutella UDP Extension for Scalable Searches), a hybrid peer-to-peer overlay builds upon the notion of ultra-peers [16]. A search is conducted by iteratively contacting different ultra-peers for their leaf nodes until a number of objects are found. These ultra-peers need not be the neighboring nodes and also the order with which ultra-peers are chosen is not specified.

In [17], author has studied the minimum delay P2P video streaming problem. For a delay sensitive application, the standard uploading bandwidth of a peer cannot be utilized to upload a piece of content until it completes the download of that content. He proposed minimum delay bound for real-time P2P systems. He has shown that the bandwidth heterogeneity amongst peers can be exploited to significantly improve the delay performance amongst peers.

In routing indices [18] based search, each node keeps information of topics and number of documents in each topic available in the neighbors. The goodness of a neighbor (compound routing index) is computed based on these statistics. The query is forwarded to neighbor with the highest compound routing index. The hop-count routing indices consist of non-cumulative number of documents at each hop. This index is computed based on number of documents and number of messages required to reach those documents. But this approach involves the cost of keeping up to date information of neighbors. In intelligent search [19], the query is forwarded to best neighbors that have responded to similar queries. The similarity is computed using cosine similarity model which is the cosine of angle between current query vector and the past answered query vector of the neighbor. For this, each node keeps a profile of answered queries for each neighbor. In adaptive probabilistic search [20], the query is forwarded to a node with the highest probability value. The probability value for a neighbor is computed based on the past query responses and current query result by the neighbor. Each node maintains a local index for each neighbor and each object. The index entry for an object and a neighbor indicates the relative probability of the neighbor being selected for querying that object. In ant based search algorithm [21], the goodness of a neighbor is judged by number of documents and path length of the neighbor. Also the goodness updation algorithm dynamically finds out optimal path for a particular query. But this approach fails in case of high churn. In directed BFS search [15], the query is forwarded to neighbors who have good statistics. This is done only for the first hop and for the rest of the hops the query is forwarded to all neighbors.

To reduce response time and bandwidth, approaches in [22, 23] specify selection of a flooding or DHT based lookup based on the popularity of the content. It is computed using a global collection algorithm. It is based on the observation that flooding is efficient for finding a popular item, but to find a rare item DHT based lookup is used.

Our technique is similar to the search technique described in [19]. But our technique differs from this in two ways, one is that the criteria to be used as performance metric in forwarding the query is given by the user and the second is that we use composite function to compute a cost that gives us the best possible route in the churn scenario.

### III. LOOKUP IN UNSTRUCTURED OVERLAYS

Lookup in unstructured P2P overlay networks happens by forwarding messages to neighbors. In figure 1, let the requesting node be A, and responding node be H i.e. say user at node A wants a video file that is stored at node H. As shown in the figure, peer A sends the requests to its neighbors B and C and they in turn forward it to their neighbors D and E. This happens until either the TTL becomes zero or item is found. When TTL reaches zero, the query is no more forwarded. Here, peer H has found a match for the query. When a peer finds a matching item, it sends a query hit message traversing the same route as taken by the query.

Techniques discussed in [14, 15] for keyword search focus on efficient and partial search but not on comprehensive search. The partial search is acceptable in case of finding a single file over the overlay. But that is not sufficient while searching for web pages, multimedia documents etc. where the information is available at multiple nodes and the requester wants to make a choice. Algorithms such as the one presented in [24] propose content addressable Publish/Subscribe service to help user find most relevant information or comprehensive information using ranked key word search. In these algorithms i.e. be it partial or comprehensive, collecting and maintaining the documents in a robust, efficient and distributed manner is a challenge. Also, in [25], issues like what should be the order of the peers to be probed while processing a query without putting much load on the peers, how to detect and prevent selfish behaviors etc., are addressed as research challenges. Efficient search can be measured in terms of the quality of service (QoS) guarantees provided by the search or lookup procedure. Here, QoS can be measured in terms of different metrics, depending on the application and a spectrum of acceptable performance along each metric. The different metrics that can be considered are response time which is a measure of bandwidth, latency, resource availability, relevance or precision of the response etc as given in [26].

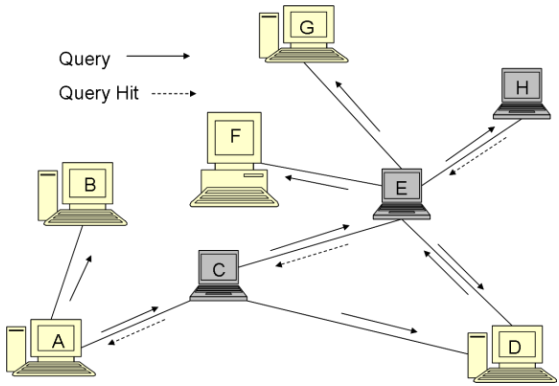

**Figure1: Lookup in unstructured P2P overlay**

## IV. SCALABLE AND ADAPTIVE LOOKUP

A peer-to-peer overlay network is formed by a set of peers. Each peer has limited knowledge about the other peers. That is every peer knows few other peers known as neighbors. Also each peer probes periodically and stores the bandwidth and link delay of the links that are connecting to its neighbors.

### A. Protocol Overview

The protocol is based on the mechanism of controlled forwarding of the queries. The control is by the user specified constraints. At each node the query is forwarded further only if the neighbor satisfies the constraints given by the user.

*User behaviour*

The user while requesting for a particular keyword, he also specifies the minimum bandwidth, maximum link delay. He expects the results that are collected from the peers in the network should satisfy these constraints. The protocol also sorts the results based on a cost metric that is calculated by combining bandwidth, link delay and past response in certain proportion.

*Query State*

In addition to the fields present in the Gnutella [3] Query message, each query consists of the minimum bandwidth, maximum latency, and composite cost.

*Node behaviour*

Each node has a data structure that stores the bandwidth and latency of the links leading to neighbors. The node probes the neighbors periodically to update the bandwidth and link delay values. The method of computing bandwidth and latency is described in part B. Also each node has a query hit history containing the addresses of the nodes from whom QueryHit message is originated and the number of files matched.

Upon receiving a query, the peer first checks up with the local database to see whether the item matching with the keyword is available. If such an item not available, the peer selects the neighbors which satisfy the condition that composite cost of that link should be less than the maximum cost calculated from the constraints specified by the user.

Composite weighted cost of each out going link is calculated. Composite cost of $i^{th}$ neighbor is calculated as follows:

Composite cost$_i$= 0.65 x (bandwidth)$_i$+ 0.20 x (delay)$_i$

All those links which have cost less than maximum cost will be selected for forwarding the query. The $i^{th}$ composite cost is added to the cost present in the query to be sent to $i^{th}$ neighbor. The cost in the message is the sum of individual link costs of route so far traveled. Figure 2 describes an example.

In figure 2, let the requesting node be A. A has set the minimum bandwidth and maximum latency as 2 Mbps and 20 msecs respectively. The cost computed by A for these constraints is 6.25. Assume that 10 Mbps bandwidth gets rating of 1 point and 100 ms link latency gets 10 points rating. This is further explained in part B. The bandwidth falls in the range of (1-2). So it gets 9 points. The latency falls into the range of (10-20). So it gets 2 points. The cost is 0.65 * 9 + 0.2 * 2 = 6.25. Node A selects only those neighbors whose link costs are less than 6.25. So, node A has selected C but not B. Node C also computes costs of the links leading to D and E. Since the cost of both the links is within the maximum cost, the query is forwarded to both D and E. Similarly at every node the composite cost of each link is computed and compared. The query has found a match at two peers F and H. They send the query hit message to the requester peer A. The replies travel through the same path as traveled by the query.

*Hit Node behaviour*

Hit node is the node which has found the one or more results for the query. This node stops forwarding the query further. It makes the Query Hit message and copies the composite cost from Query message to Query Hit message. Query Hit message also includes the number of results found. The query hit message is sent over the same route through which the query has come. In Figure2, F and H are the hit nodes.

*Requester Node behaviour*

Requester node is the one which is performing the lookup. Upon receiving a query hit, the requester node adds the address of the node and the number of files present in the Query Hit message to its query hit history. It also updates the past response of the node. The detail of how the past response is updated is explained in part B. The composite cost is updated in the following manner.

(Composite cost)$_i$= (composite cost)$_i$ + 0.15 * (past response)$_i$

The composite cost of $i^{th}$ query hit message is updated by adding the past response of the query hit node with 15% weight.

After receiving the query hit messages, the node sorts them by the composite cost. The user can chose the first displayed result to get the file.

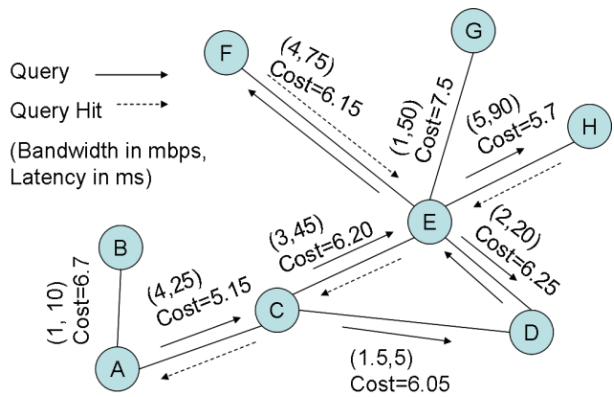

**Figure2: QoS based Lookup in unstructured P2P overlay**

After receiving the Query Hit messages over a period of time, the node sorts them by the updated composite cost. The lowest cost is ranked as 1 and highest cost is ranked last. If there is a tie amongst the composite costs received, tie is resolved by considering the number of files that node has returned in the current response.

In figure 2, requesting node A receives query hit messages from nodes F and H. The cost of the route to F and H is 17.50 and 17.05 respectively. Assuming that F and H have past responses as 5 and 4 respectively and they have returned 14 and 6 documents in the current query, their past responses will be updated to 4.6 and 5 respectively. The updated composite costs of F and H will be 18.19 and 17.80. So, peer H will be given rank 1. Here we assumed that maximum number of files that a node can return is 50.

*B. Composite Function*

The user is interested in several factors such as number of results, proximity to expected content, maximum bandwidth and minimum latency of the route to download the file etc. Here we try to optimize the bandwidth and link latency for the route. Whenever query is sent, the expected QueryHit will be the one that can provide an optimum path i.e., the maximum speed of transfer, low time delay and a non-corrupted file. But maximum preference is given for the speed of the transfer and then latency. Here latency includes all kinds of delays. In this protocol, a weight of 65% to the transfer speed and 20% to latency and the remaining 15% to the past response (popularity of the peer) is given. This weightage can be changed according to the requirements of the overlay network.

*Bandwidth*

The bandwidth available with a link associated with a node is calculated by sending a packet of very small size to its opposite node. After receiving the query hit message the ratio of packet size sent to the round trip time gives the bandwidth available with that particular link. However, we have assumed other delays associated in the packet transfer as negligible in calculating the bandwidth available over a link. This process is done for all the neighbors of a particular node.

*Latency*

Link latency is computed as the propagation delay over the link that is directly proportional to the distance of the link between two peers. For simplification, we do not consider any queuing delay and processing delay while computing link latency. However, these delays will not have any adverse effect in our Protocol.

```
Algorithm 1:
/* Pseudo code for a node which receives a query*/
ProcessQuery(Q){
N: set of all neighbors
bw: Array of bandwidths of links of  neighbors
ll: Array of link latencies links of neighbors
Q: Query that has come from another node
SN: Node that has sent the query
MC: Maximum cost as per user requirements

/* avoids loop */
if Q.message id found in local cache then
        drop the query, exit
end if
if Q.keyword matches files in local db then
        make Query Hit Message
        send Query Hit Message to SN
else if Q.hopcount = 0 then
        drop the query
    else
        store the Q.MessageId in cache
        for each neighbour in N
          if N <> SN then
          if computeCompositeCost (neighbor) <= MC
        then
            CQ=copy of Q
            CQ.hopcount=CQ.hopcount-1
            CQ.cost=CQ.cost+cost of neighbor
            forward CQ to neighbor
          end if
          end if
        end for
end if }
/* Function for computing cost of link to neighbor*/
ComputeCompositeCost(neighbor){
        BW: bandwidth of link to neighbor
        LL: link latency
        NBW: normalized bandwidth
        NLL: normalized link latency
        MAXBW: Maximum bandwidth possible in the network
        MAXLL: Maximum link latency possible in the network
        /* The ranges are divided as follows
           (0 , MAXBW/10] = rating 10
           (1 * MAXBW/10, 2* MAXBW/10] = rating 9
           ....
           (9 * MAXBW/10, MAXBW] = rating 1
        The ideal conditions are given lowest ratings
        Similarly the latency is also divided into ranges
        with ratings
           (0 , MAXLL/10] = rating 1
           (1 * MAXLL/10, 2* MAXLL/10] = rating 2
           ....
           (9 * MAXLL/10, MAXLL] = rating 10*/
        NBW = rating of BW
        NLL = rating of LL

        return (0.65 * NBW + 0.20 * NLL)
}
/* pseudo code for requester node */
ReceiveQueryHit(){
QH: Query hit message
N: Node form which query hit originated
i: rank
for each query hit QH received from node N
        if N not found in local history cache then
                store the address of N in local history cache
        end if
        retrieve past response for N
        past response= 0.8 * past response + 0.2 * rating for
        QH.number of documents matched
        QH.cost=QH.cost + 0.15 * past response of N
        save past response in history cache
end for
i=1
for each query hit in {QH sorted by QH.cost,QH.no of files in
ascending order}
        assign rank i
end for
        display results for user with ranks
}
```

*Past reposne*

Past response of a node indicates the reliability of the node. It is calculated for every node returning a query hit message.

The main problem that has to be looked into is the normalization of various parameters that were being assigned some weightages in the calculation of the composite function. This is solved by scaling all the parameters on a scale of 10. As we are seeing for the minimization of the composite function the ideal conditions must be given the lowest rating and the worst conditions the highest.

Assuming the maximum bandwidth available in the network to be B and as the minimum bandwidth can go until 0 the bandwidth available with a particular link can be given the rating. Depending upon the maximum and minimum bandwidths available, a rating of 1 means that the bandwidth available is in the range (9B/10 – B] and a rating of 10 implies that the bandwidth available is in between (0 - B/10].

Now the scaling of latency values on a scale of 10 is done using the same process as above and the maximum latency L is an assumed value throughout the global network. Here the latency in the range (9L/10 - L] is given rating of 10 because that is the ideal condition.

Past response is calculated as the 80% of the old past response and 20% of the normalized rating for the number of files returned in the current hit message. To compute the past response, we used the following formula:

(Past response)$_i$ = 0.8 * (past response)$_i$ + 0.2 * (rating points for the number of files returned in the current query hit message)$_i$

The scaling of response of a node on a scale of 10 is done in the same as mentioned for bandwidth. If the maximum response is P, then if a node returns number of files falling in the range (9P/10 – P], it is given rating of 1 and other ranges are scaled accordingly.

### C. Algorithms

The pseudo code for processing the query received from a neighbor, computing composite cost and processing the hits received from peers is given in Algorithm 1.

### D. Adaptive nature of our approach

Our approach finds the most preferable overlay route under the constantly changing link bandwidths and delays. It adapts to the dynamically changing network parameters by choosing the best neighbors at every hop. Also the algorithm adapts to the high churn (node joining and leaving) scenario mostly found in the P2P overlay networks. This is possible because the approach always finds the best neighbors based on the bandwidth and link latency but not on the content that is hosted by the peer. Our protocol also handles the link or node failures which could be the result of an earlier connected peer leaving the overlay network. In this situation the protocol automatically recalculates the better available route for the next set of queries that arrive at the existing peers. The QoS parameters that the user is able to specify while requesting for the object could be based on his own experiences with the access network to which he is connected and the resource available at the end system or it could even be on the basis of his experiences with the peer-to-peer overlay download over few days or months or years. A peer that was down because of some reasons (possibly crash or maintenance reasons) when rejoins, our algorithm dynamically integrates it into the existing set of overlay peers.

## V. SIMULATION RESULTS

This section shows our simulation results. The objective of the simulation is to show that the proposed protocol indeed delivers the expected results and consumes low bandwidth of the network and hence can accommodate more number of peers without degrading performance. The simulated network consists of 1000 nodes and 50 different objects but spread randomly across the network. The degree of node varies from 3 to 12 with average 6. The objects are distributed randomly across the peers. Each peer has maximum of 15 objects. The bandwidth and link latency is randomly assigned to each link. We compared our approach with the flooding technique which is used by most currently operating unstructured overlays. The TTL limit is varied from 1 to 5. The bandwidth of the links is randomly changed during simulation to reflect the dynamic nature of the network congestion. It is changed after every query. We plotted the results as shown in figure 3 and figure 4.

As we see in figure 3, as the TTL increases the message overhead produced by the QoS based search becomes insignificant when compared to flooding. But when the hop count reaches 4, there is an exponential increase in the message overhead for the flooding approach. When the hop count is 5, the message count raises up by a factor of several thousands. But in our adaptive search approach, increase in message overhead is very slow with respect to hop count. This is natural out come of the QoS constraints. The message overhead directly affects the bandwidth consumed in the network. The graph in figure 4 shows that the magnitude of results returned by the two approaches. When hop count is 1, the results returned by the both approaches are zero. The results returned by flooding increases rapidly with respect to the hop count. As shown in figure 5, flooding approach returns huge number of unwanted hits. These results do not satisfy the user requirements. They unnecessarily consume the bandwidth doing nothing good for the user and reducing the scalability.

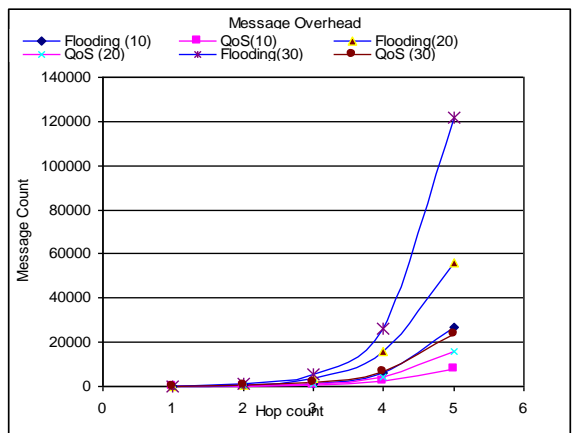

**Figure 3: Performance of QoS based search** (Numbers in the brackets indicate number of queries)

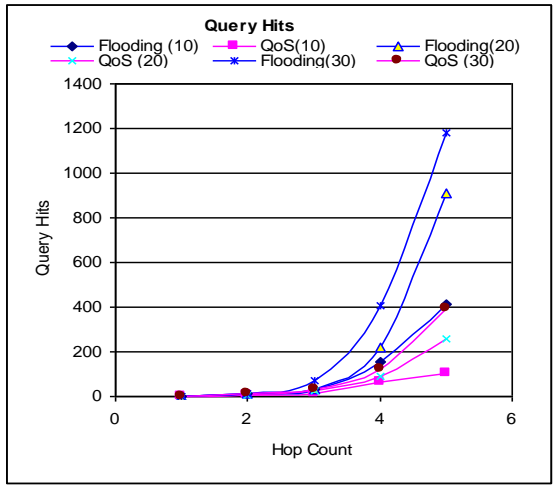

**Figure 4: Query hits of QoS based search**
(Numbers in the brackets indicate number of queries)

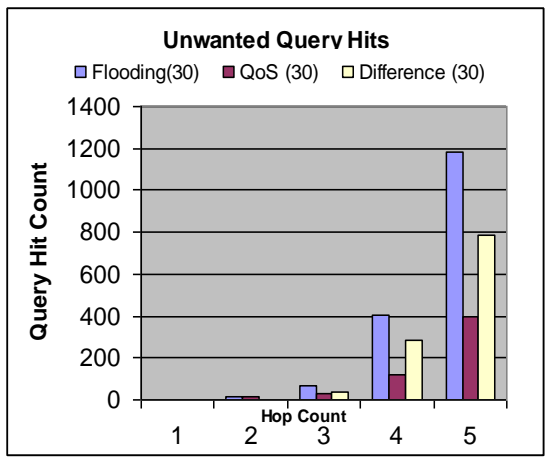

**Figure 5: Unwanted query hits**
(Numbers in the brackets indicate number of queries)

## VI. CONCLUSION

In this paper we presented a QoS based adaptive heuristic search protocol for unstructured peer to peer overlay networks. The objective of this heuristics is to find items in the routes that satisfy the user given constraints. The approach also aims at reducing the bandwidth consumption of the network. We quantify the performance of our approach in terms of number of hits and message overhead in the overlay network. Our adaptive heuristics performs better over flooding technique which is currently being used predominantly in most of the unstructured P2P overlays. For small TTL values, both flooding and our approach do not differentiate much. However, for moderate to large TTL values, our adaptive protocol improves performance of the overlay lookup. Our protocol is scalable because it judiciously or optimally uses the bandwidth available over the overlay links. Our approach also handles Churns (high rate of peer joins and leaves) efficiently by choosing the right neighbors who have low latency and less congestion amongst the available neighbors. However, in literature researchers have selected nearest neighbor in the churn scenario which may not be better always from performance metric considering the QoS angle. We plan to extend our work by building a testbed for unstructured P2P overlay in our advanced network research laboratory running a modified version of open source software like bittorrent with our adaptive lookup being used to improve the download speed of multimedia contents in a large peer base.